# Thin Lead Titanate films grown by Molecular Beam Epitaxy

GIJSBERT RISPENS* AND BEATRIZ NOHEDA

Zernike Institute for Advanced Materials, University of Groningen,
Nijenborgh 4, 9747AG Groningen, The Netherlands

The growth of atomically-flat thin films of ferroelectric $PbTiO_3$ on $SrTiO_3$ substrates, using molecular beam epitaxy, is reported. The main issue in the growth of these materials is the high volatility of lead. This can be largely overcome by using PbO, instead of Pb, as a source and by using atomic oxygen during growth. The continuous decrease of the out-of-plane lattice parameter with increasing temperature in the investigated range, indicates that $PbTiO_3$ is still ferroelectric at the growth temperature ($T_g$= 600°C), which agrees with the theoretical prediction of $T_C$= 765°C (compared to $T_C^{bulk}$= 490°C) for the present mismatch strain values.

## INTRODUCTION

Lead titanate is a classical ferroelectric material that continues to raise interest. On the one hand, it is a model system with one of the largest polarizations and c/a values measured so far. On the other hand, it is the basis for other technologically important ferroelectrics such as $Pb(Zr_{1-x}Ti_x)O_3$ (PZT) or $Pb(Mg_{1/3}Nb_{2/3})O_3$-$PbTiO_3$ (PMN-PT). Especially, thin films of $PbTiO_3$ are widely studied. Of significant research interests are the influence of epitaxial strain and depolarizing fields on the ferroelectric properties and the formation of periodic domain structures [1, 2, 3]. Also the effect of thickness on the ferroelectric properties has been previously studied [4,5,6]. An important issue in these ultra-thin films is leakage. This leakage is partly due to defects such as cation and oxygen vacancies [7,8].

The growth of lead titanate thin films has been reported by many groups using various methods, such as pulsed laser deposition (PLD) [3], off-axis RF sputtering [4] and MOCVD [2]. Although the growth of $PbTiO_3$ thin films using molecular beam epitaxy (MBE) has been reported previously by one group [10,11], there is little known about the actual properties of $PbTiO_3$ films grown by this technique. Here we report the growth of $PbTiO_3$ thin films using MBE. The final goal of this effort is to compare the properties of these thin films to those of films grown using other, more common methods, especially PLD, and through this, learn about the growth-specific defects and their influence on the physical properties.

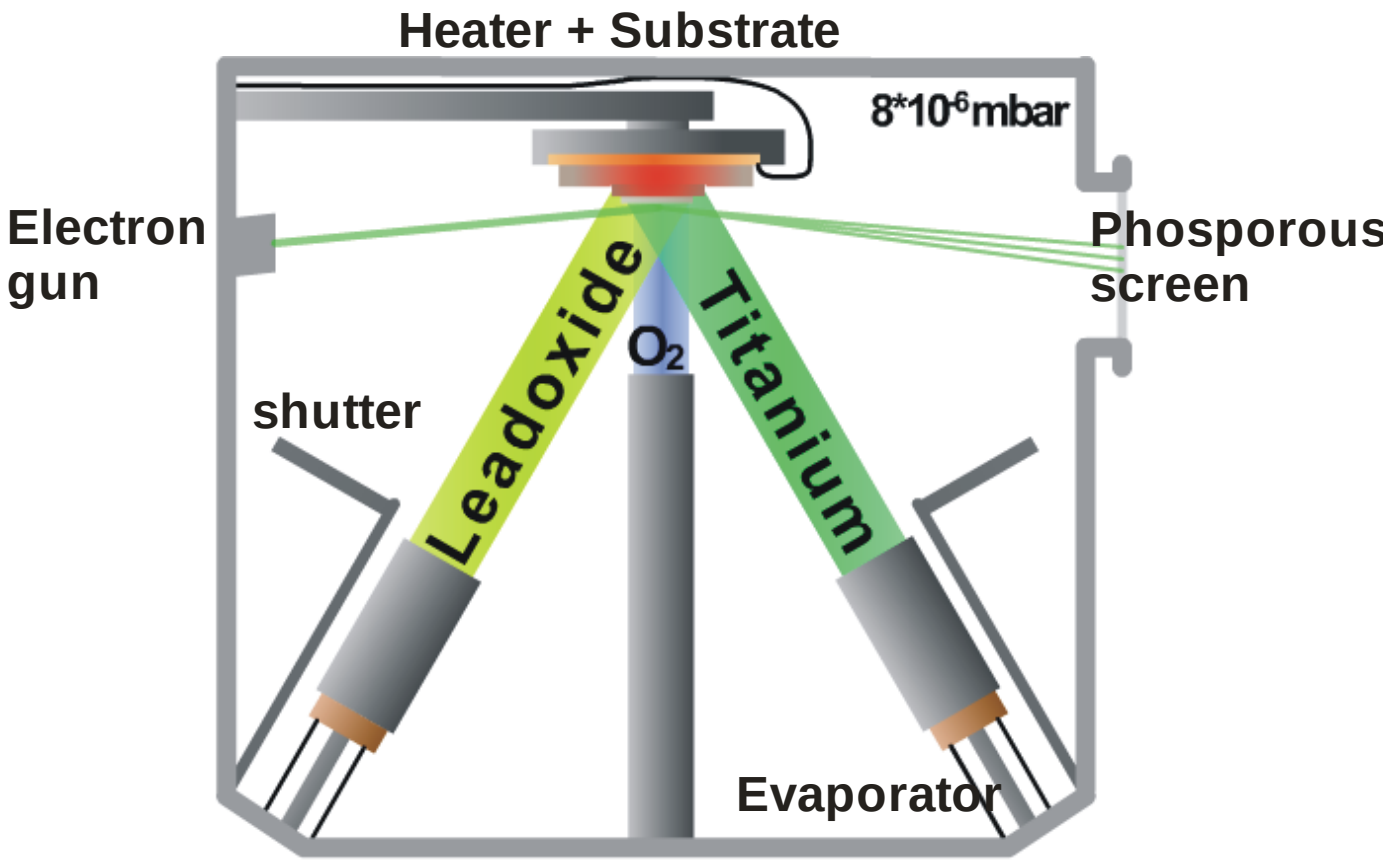

Figure 1: Schematic drawing of the setup used for the MBE growth of $PbTiO_3$. The oxygen source can be, either molecular oxygen, ozone, or atomic oxygen from a RF plasma source

## EXPERIMENTAL

Epitaxial thin films of $PbTiO_3$ were grown using Molecular Beam Epitaxy (MBE). The samples were grown in a modified VG MBE chamber, a sketch of which is shown on Figure 1. Titanium was evaporated from a Omicron Focus EFM 4 electron beam evaporation source. The flux was measured and controlled using the flux monitor of the evaporator. Lead oxide (PbO) was evaporated from a Luxel Radak thermal evaporation cell. Atomic oxygen was supplied by a Oxford Scientific ECR plasma source fitted with a quartz delivery tube to ensure a high flux a the substrate surface (see figure 1). The films were grown on $SrTiO_3$ substrates. The $SrTiO_3$ substrates were treated chemically and thermally to ensure a single termination and atomic steps on the surface [11]. The substrates were mounted on stainless steel sample holders using platinum paste for good thermal contact and spot-welded tantalum strips. The substrates were heated using a filament. Prior to the growth, the fluxes were calibrated using a quartz crystal monitor. The chamber is equipped with a Staib electron gun and a phosphorous screen for reflective high energy electron diffraction (RHEED). The electron energy used for RHEED was 15 KeV.

An important issue in the growth of thin films is control of the stoichiometry. In the case of MBE growth of complex oxides this is particularly a problem because one has to ensure that stoichiometric amounts of two or more metals arrive at the substrate surface. Therefore in-situ flux control using atomic absorption is widely used in oxide MBE systems. Another possible method is to use the evolution of the RHEED intensity to time the shuttering of the beams. RHEED can be used to follow the growth by monitoring the intensity of the electron diffraction spots. The intensity of the spots is at a maximum when a layer is filled and has a minimum when a layer is half filled. So if the RHEED intensity is followed during growth, oscillations can be observed if the material grows layer-by-layer (see figure 2). If RHEED is used to control the shuttering of the beams, full atomic monolayers of the simple oxides can be stacked to form the perovskite cell. This method only works if the simple oxides can be grown layer by layer.

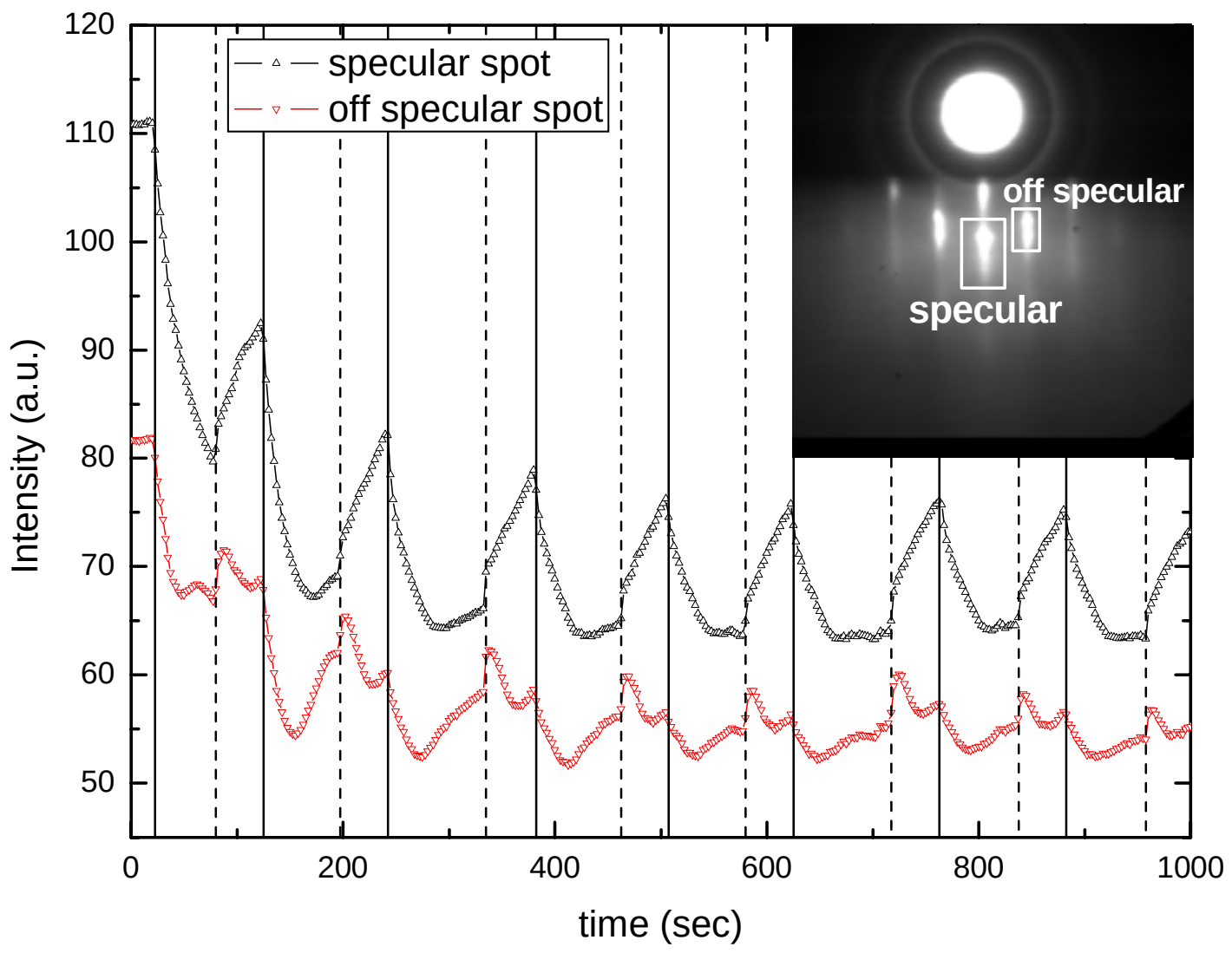

Figure 2: RHEED intensity oscillations during the first 1000 seconds of the growth of $PbTiO_3$ on $SrTiO_3$. The dashed lines indicate the moment the Ti shutter was opened; the shutter was closed again at the solid line. The maximum intensity corresponds to a complete layer. The inset shows a typical RHEED pattern as observed during the growth of the film.

The main issue in the growth of $PbTiO_3$ films is the volatility of the lead. Moreover, the $PbTiO_3$ perovskite structure allows for large amounts of Pb vacancies[9]. To maintain charge neutrality Pb vacancies often come together with oxygen vacancies. To reduce this problem, we have used PbO as a lead source, instead of metallic Pb. This serves two purposes: First of all, the vapour pressure of PbO is slightly lower than that of Pb, so a smaller re-evaporation from the substrate surface is expected. A second advantage of PbO is that the kinetic and thermodynamic barriers for oxidation of lead are removed. Despite its advantages the sticking coefficient of PbO is still very low, so a large excess of PbO has to be supplied during growth. While co-evaporation of Ti and PbO almost invariably resulted in lead-deficient perovskite films, sequential deposition of PbO and $TiO_2$ monolayers, largely improved the films quality. We tried various oxidation methods using molecular oxygen, distilled ozone and atomic oxygen as oxygen sources. The use of atomic oxygen provided the best results.

The best films were obtained using a method similar to the one described by Theis and Schlom [10]. With this method, we use the volatility of PbO in our advantage by relying on the fact that all PbO will evaporate back into the vacuum unless it is stabilized into the perovskite structure. Thus, PbO is supplied to the substrate continuously while the titanium flux is supplied in doses of one monolayer. The titanium flux could be controlled in-situ using the flux monitor of the evaporator. The fact that the PbO is open continuously minimizes re-evaporation of PbO from the film. This method also depends heavily on the flux and oxidizing power of the supplied oxygen. The range of substrate temperatures available for $PbTiO_3$ growth is limited on the low end by the need for sufficient diffusion to form smooth films and it is limited by the loss of lead oxide on the high end. The best results so far were found at a growth temperature close to $600^oC$.

The grown films were characterized using x-ray diffraction and atomic force microscopy (AFM). AFM images were recorded using a Digital Instruments Nanoscope III multimode microscope. X-ray patterns were collected using a Philips X'pert MRD diffractometer. The temperature dependence of the lattice parameters was measured using an Anton Paar DSH900 hot stage fit to our diffractometer.

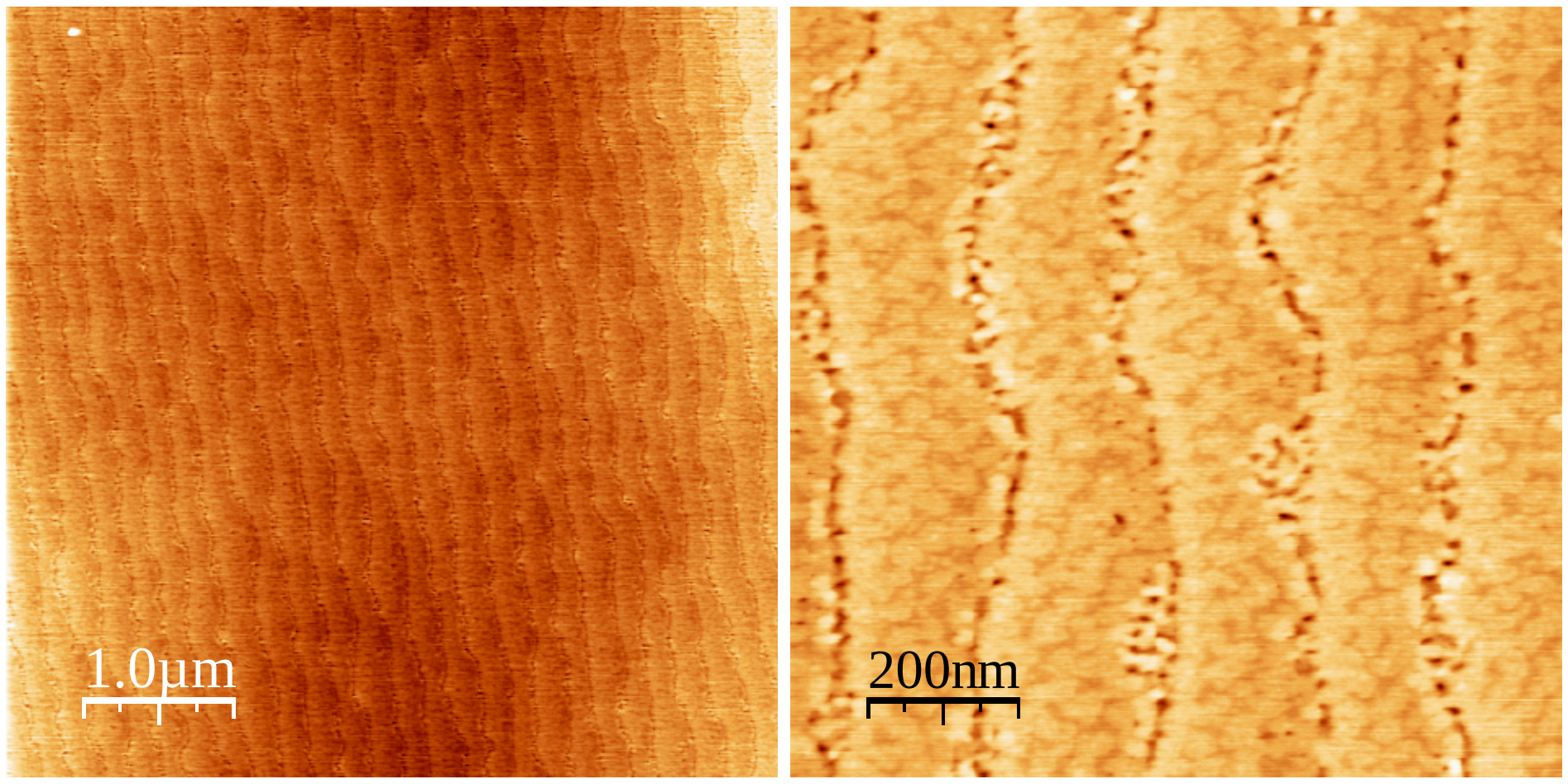

Figure 3: AFM images of the surface of a 7 nm thick $PbTiO_3$ film on $SrTiO_3$. The atomic steps of the substrate are still visible. Some nucleation of a new layer is also visible. The RMS roughness is ~0.3 nm, also consistent with the layer-by-layer growth mode. [12]. The contrast at the step edges and the meandering of the terraces originate in an imperfect treatment of the substrate.

## RESULTS AND DISCUSSION

In the AFM image shown in Figure 3, the morphology of a 7 nm thick film can be observed. The terraces of the substrate are still visible after deposition, indicating the 2-dimensional growth mode of the $PbTiO_3$ film (as opposed to island-like mode), in agreement with the RHEED oscillations of Figure 2 . The terraces of the substrate are due to the miscut between the surface and the crystalline planes. The roughness of the films is typically below one unit cell.

X-ray reflectivity scans are used to verify the thickness of the grown films as well as the critical angle of total reflection. This critical angle is a rough measure of the density of electrons in the film and thus can give some information about the stoichiometry, where a low electron density is assigned to the existence of lead vacancies (likely associated to oxygen vacancies). To extract this data, the reflectivity scans were simulated [13]. As shown in Figure 4, the fit is very good and there is no significant deviation from the nominal electron density of $PbTiO_3$. The thickness, of 6 nm in this particular case, is in agreement with the number of oscillations observed in RHEED.

From the linear 2θ-ω scans (taken with an x-ray wavelength of 1.5405 Å), as the one shown in Figure 5, and the fit to the crystal truncation rods, the out of plane lattice parameter (*c*) of the films is extracted. The out-of-plane lattice parameter was found to be about $c$=4.12 Å, which corresponds to a $c/a$ ratio (tetragonality) of 1.055 at room temperature. Although this value is slightly smaller than the bulk value, it is large compared to that of single domain films of the same thickness[5] The suppression of the *c/a* ratio at small thickness is believed to be due to the presence of depolarization fields, which originate from uncompensated surface charges. The high *c/a* ratio in our as-grown films suggests that there is good screening of the surface polarization charges. This could be attained by means of charged adsorbates during growth and subsequent cooling, or due to the influence of the RHEED electron beam, provided we used a metallic bottom electrode. Since this is not our case, we can only expect this to happen if large amounts of charged impurities were present at the interface. A more likely mechanism, in our case, to compensate net charges at the interfaces, is the formation of 180$^o$ stripe domains. Direct evidence of periodic stripe domains, as those reported in other films [2, 3], is not observed in the x-ray diffraction patterns of these films, which could be due the lack of coherence of the stripes

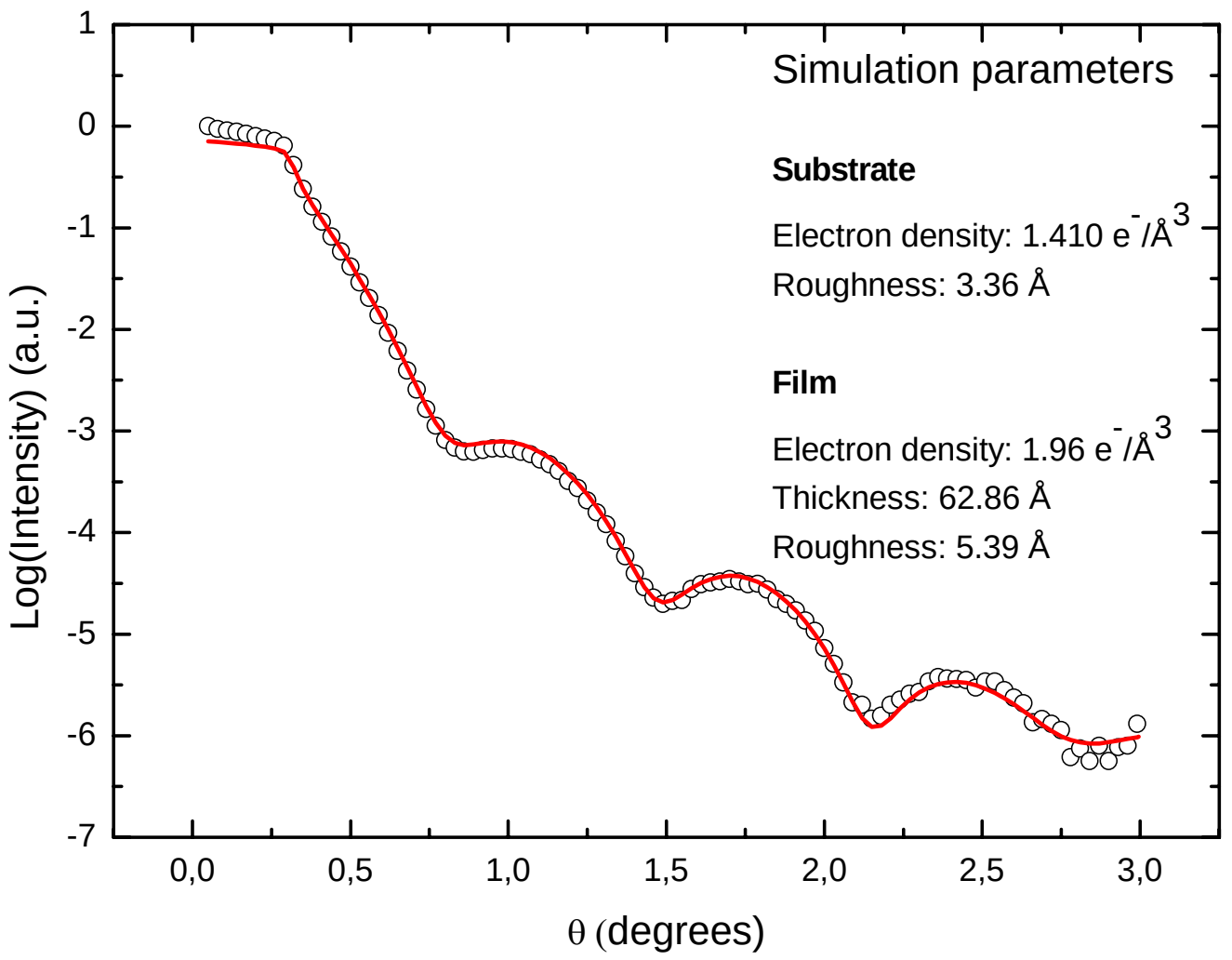

Figure 4: X-ray reflectivity scan of a 6 nm thick $PbTiO_3$ film on $SrTiO_3$. The symbols are the measured data, the line is a simulation. From the simulation, the electron density was found to be the one expected for stoichiometric lead titanate.

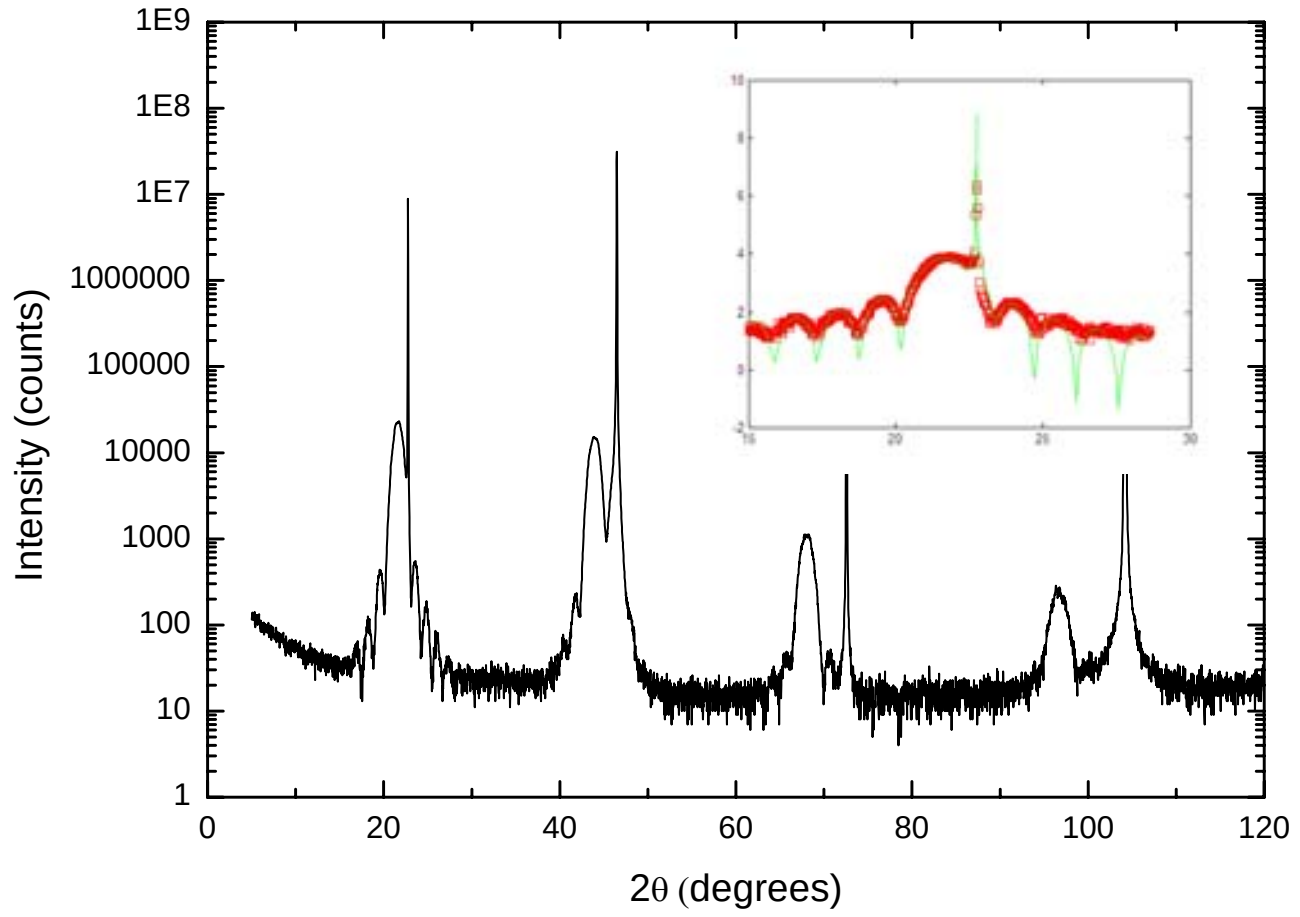

Figure 5: Specular 2θ-ω scan of a 7 nm thick $PbTiO_3$ film on $SrTiO_3$. The c lattice parameter from this scan is 4.12Å. The fact that the finite size oscillations are visible is an indication of a good film quality. The inset shows the fit of the 001 reflection used to extract the out-of-plane lattice parameter of the film.

A reciprocal space map around the off-specular (103) reflection can be found in Figure 6. Such reciprocal space maps were used to confirm the epitaxy of the films. It can be seen that both the film and the substrate share the same in-plane lattice parameter and thus the films are epitaxial.

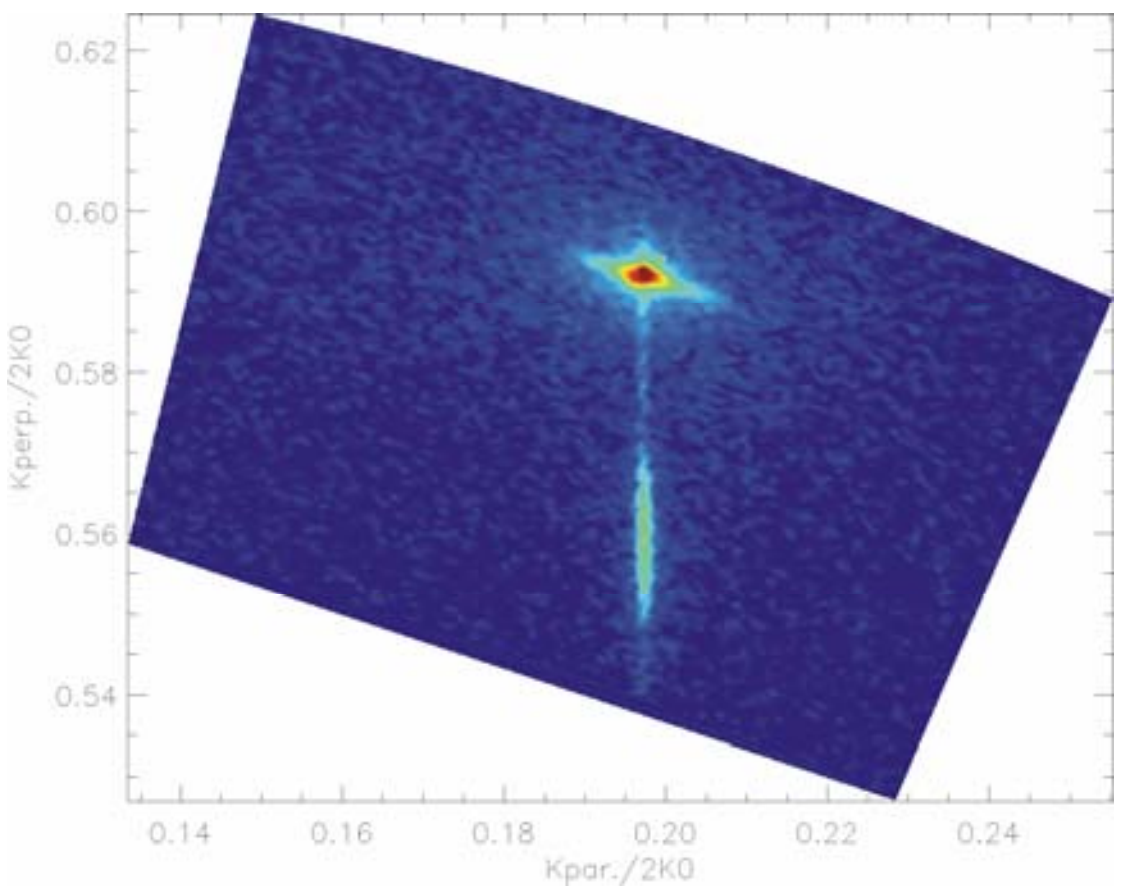

Figure 6: Reciprocal Space Map around the (1 0 3) reflection. Both the substrate peak (top) and the film peak (bottom) share the same $k_{parallel}$ and thus the same in-plane lattice parameter. The axis are in unit of $4\pi/\lambda$, with $\lambda = 1.5405$ Å.

The temperature dependence of the lattice parameters of one of the films is shown in figure 7. The out-of-plane lattice parameter decreases with increasing temperature, as one would expect for a ferroelectric approaching the transition to the paraelectric phase, whereas the in-plane lattice parameter follows the thermal expansion of the substrate. At lower temperatures, the data are in good agreement with those expected for the lattice parameters of bulk $PbTiO_3$,[14] strained in-plane by $SrTiO_3$, as calculated simply using the Poisson ratio. The Poisson ratio (ν) relates the out-of-plane strain ($\varepsilon_{out-of-plane}$) to the in-plane strain ($\varepsilon_{in-plane}$) as

$$\frac{\varepsilon_{out-of-plane}}{\varepsilon_{in-plane}} = -\frac{2\upsilon}{1-\upsilon}$$

In the case of an epitaxial film, the in-plane strain is determined by the mismatch between the film and the substrate, so the out-of-plane strain and thus the out-of-plane lattice parameter can be calculated provided ν is known. Here we use ν = 0.3125, as calculated from the elastic compliances of bulk $PbTiO_3$[15]. The agreement between the measured lattice parameters (solid symbols) and those expected from strained $PbTiO_3$ at temperatures sufficiently far from $T_c$ supports the idea that the lattice parameter is not suppressed by depolarization effects.

At higher temperatures (corresponding to larger misfit strain values), the behavior is not so clear and the previous approach produces a very small unit cell volume. As mentioned above, the dashed line in Figure 7 considers pure elastic effects and does not take into account neither the increment of the transition temperature expected in these films due to electrostrictive effects [15], nor the changes of the order of the transition (from discontinuous to continuous) due to epitaxial strain[15,16], nor the influence of the 180$^o$ domains in the temperature evolution of the lattice parameters. When these effects are included in a phenomenological Landau-like approach[15,16], and at the limit of thick films, the solid line in Fig. 7 is predicted. The observed c-values are significantly lower than the theoretical ones, as expected for such thin films. At T*=450-500$^o$C, there is an anomaly and a change of bevaviour in the experimental temperature dependence, similar (although more significant) to the small predicted dip at ~500$^o$C [16]. Above this temperature, the experimental values lie in between the two theoretical curves. Loss of Pb or PbO from the film cannot be discarded above T~ 600$^o$C and this sets and upper limit to our measurements, preventing us for reaching the paraelectric phase.

Upon subsequent cooling, the *c* lattice parameter appears considerably smaller (open symbols in Fig. 7) and the films do not revert to their original state after cooling. Although the good agreement between the heating and cooling lattice parameters observed at 600$^o$C seem to argue against it, this discrepancy could be due to the creation of lead vacancies during heating in air. Another possible explanation is that the domain structure has been altered so that stronger depolarization fields are present after the first heating run, reducing the c parameter.

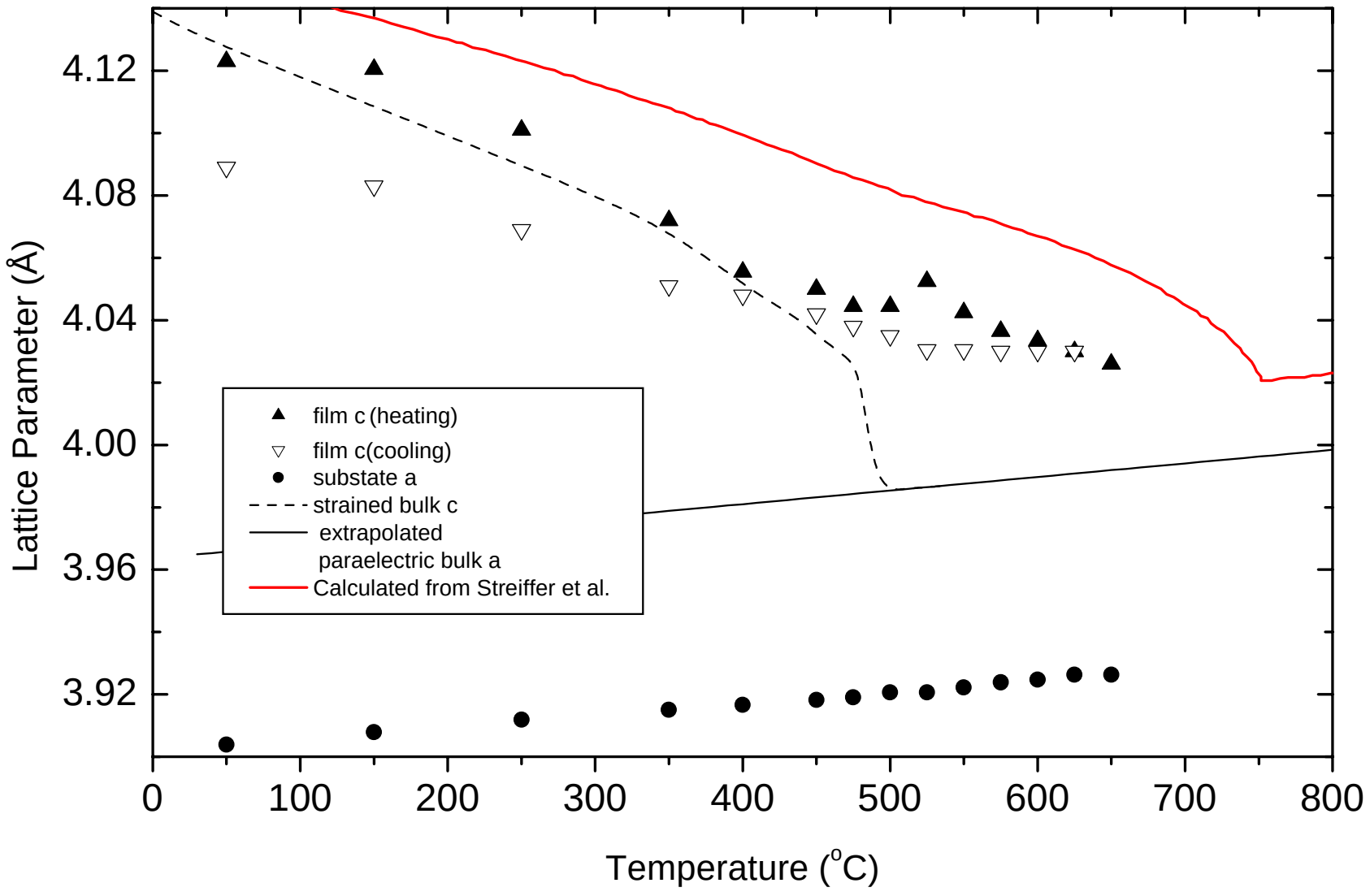

Figure 7: A) Temperature dependence of the lattice parameters of a 7 nm thick $PbTiO_3$ film on a $SrTiO_3$ substrate. Upon heating (closed up triangles), and for temperatures below T~ 470°C, the *c* lattice parameter is in reasonable agreement with the value found for bulk $PbTiO_3$ strained to fit the in-plane lattice parameter of the susbtrate and using the reported Poisson ratio of 0.312 (dashed line). The straight line is the *a* lattice parameter expected for strained paraelectric $PbTiO_3$ exptrapolated to lower temperatures. The *c* parameters during subsequent cooling are represented by open down triangles. The red solid surve is the prediction for thick films, taken from ref. [16]

## CONCLUSIONS

Thin films of $PbTiO_3$ have been grown by MBE. The main issue in the growth process is the high volatility of lead. The use of an adsorption-controlled growth process in combination with the use of lead oxide, instead of lead, helped to grow atomically-flat layer-by-layer films with thickness ranging between 5 and 25 nm. The oxidation power of the supplied atomic oxygen has an important influence on the quality of the grown films: The higher the oxidation power, the lower the volatility of lead.

The structural quality of the films is excellent and they are completely epitaxial. The large out-of-plane lattice parameter of the as-grown films points to a good screening of the surface polarization charges, most likely due to the formation of stripe domains. The temperature dependence of the lattice parameters below 600°C shows the expected trend toward the transition temperature. We argue that a change in the 180° domain structure upon heating, involving changes in surface charge compensation, could be responsible for the observed difference in c lattice parameters during heating and subsequent cooling.

At higher temperatures (for which there are larger mismatch strain values between film and substrate), electrostiction and the change in the transition order are responsible for the large discrepancy between the pure elastic model and the observed behavior. Further characterization of the MBE grown films for different thicknesses will help understanding these results. Characterization of the functional properties is also in progress in order to compare them with those of more commonly available PLD films. This will allow us to gain insight on the intrinsic versus growth-specific defects in this tetragonal ferroelectric with one of the largest order parameters known so far.

AKNOWLEDGEMENTS

We would like to thank Henk Bruinenberg for his valuable technical assistance and Gustau Catalan, Szilard Csiszar, Tjipke Hibma and Ard Vlooswijk for useful discussions. This work is part of the research programme of the 'Stichting voor Fundamenteel Onderzoek der Materie (FOM)', which is financially supported by the 'Nederlandse Organisatie voor Wetenschappelijk Onderzoek (NWO)'.

* e-mail: g.rispens@rug.nl

.